\documentclass[conference]{IEEEtran}
\IEEEoverridecommandlockouts
  
\usepackage{cite}
\usepackage{amsmath,amssymb,amsfonts}
\usepackage{algpseudocode}
\usepackage{multirow}
\usepackage{array}
\usepackage{adjustbox}
\usepackage{booktabs}
\usepackage{graphicx}
\usepackage{textcomp}
\usepackage{xcolor}
\usepackage{subcaption}
\usepackage{caption}
\usepackage{tabularx}
\usepackage{float}
\usepackage{url}
\usepackage{algorithm}
\usepackage{makecell}
\usepackage{enumitem}
\usepackage{hyperref}

\def\BibTeX{{\rm B\kern-.05em{\sc i\kern-.025em b}\kern-.08em
    T\kern-.1667em\lower.7ex\hbox{E}\kern-.125emX}}
\begin{document}

\title{Visual Lifelog Retrieval through Captioning-Enhanced Interpretation
}

\makeatletter
\newcommand{\linebreakand}{%
  \end{@IEEEauthorhalign}
  \hfill\mbox{}\par
  \mbox{}\hfill\begin{@IEEEauthorhalign}
}
\makeatother

\author{
\IEEEauthorblockN{Yu-Fei Shih}
\IEEEauthorblockA{\textit{Department of Computer Science} \\ \textit{and Information Engineering} \\
\textit{National Taiwan University}\\
Taipei, Taiwan \\
yfshih@nlg.csie.ntu.edu.tw}
\and
\IEEEauthorblockN{An-Zi Yen}
\IEEEauthorblockA{\textit{Department of Computer Science} \\
\textit{National Yang Ming Chiao Tung University}\\
Hsinchu, Taiwan \\
azyen@nycu.edu.tw}
\linebreakand
\IEEEauthorblockN{Hen-Hsen Huang}
\IEEEauthorblockA{\textit{Institute of Information Science} \\
\textit{Academia Sinica}\\
Taipei, Taiwan \\
hhhuang@iis.sinica.edu.tw}
\and
\IEEEauthorblockN{Hsin-Hsi Chen}
\IEEEauthorblockA{\textit{Department of Computer Science} \\ \textit{and Information Engineering} \\
\textit{AI Research Center (AINTU)}\\
\textit{National Taiwan University}\\
Taipei, Taiwan \\
hhchen@ntu.edu.tw}
}

\maketitle

\begin{abstract}
People often struggle to remember specific details of past experiences, which can lead to the need to revisit these memories. Consequently, lifelog retrieval has emerged as a crucial application.
Various studies have explored methods to facilitate rapid access to personal lifelogs for memory recall assistance. 
In this paper, we propose a Captioning-Integrated Visual Lifelog (CIVIL) Retrieval System for extracting specific images from a user's visual lifelog based on textual queries. 
Unlike traditional embedding-based methods, our system first generates captions for visual lifelogs and then utilizes a text embedding model to project both the captions and user queries into a shared vector space.
Visual lifelogs, captured through wearable cameras, provide a first-person viewpoint, necessitating the interpretation of the activities of the individual behind the camera rather than merely describing the scene. To address this, we introduce three distinct approaches: the single caption method, the collective caption method, and the merged caption method, each designed to interpret the life experiences of lifeloggers.
Experimental results show that our method effectively describes first-person visual images, enhancing the outcomes of lifelog retrieval. Furthermore, we construct a textual dataset that converts visual lifelogs into captions, thereby reconstructing personal life experiences. 

\end{abstract}


\section{Introduction}\label{sec:intro}
Recent advancements in smartphones, smartwatches, and wearable cameras have led to a significant increase in personal data generation, including images, geographical information, and physiological signals. This data offers insights into daily routines and applications in areas like diet management~\cite{6563978} and memory recall~\cite{Yen_Huang_Chen_2021}. 
Therefore, an effective retrieval system to organize and access specific instances is crucial.

This paper addresses the challenge of visual lifelog retrieval, which aims to retrieve specific moment images from a lifelogger's dataset using textual queries. For example, we expect the system to retrieve the left image in Figure~\ref{fig:compare} given the query ``The moment that the individual is in the airport terminal''. The key challenges in this task include:

\begin{enumerate}
  \item \textbf{Semantic Gap:} A significant semantic gap exists between textual queries and lifelog images~\cite{cad7ae14c7f047749df68f0a7f74a70b,chu2021vidlife}. 
  For instance, to retrieve images of ``driving to the beach,'' the system must go beyond object recognition and infer the activity from visual cues like a steering wheel and ocean views. 
  
  \item \textbf{Unique Characteristics of Visual Lifelogs:} 
  Unlike most activity recognition tasks that use third-person images from standard datasets, visual lifelogs are typically captured from a first-person perspective, requiring the inference of actions behind the camera from the visible scene.
  
  \item \textbf{Identifying Lifelogger Activities from Sequential Images:} 
  Analyzing images in sequence provides a fuller understanding of the user’s activities, locations, and context. For example, as shown in Figure~\ref{fig:compare}, the location in the right image may seem ambiguous but can be identified as an airport lounge with the information of the left image.
\end{enumerate}

\begin{figure}[t]
  \centering
  \includegraphics[width=0.6\linewidth]{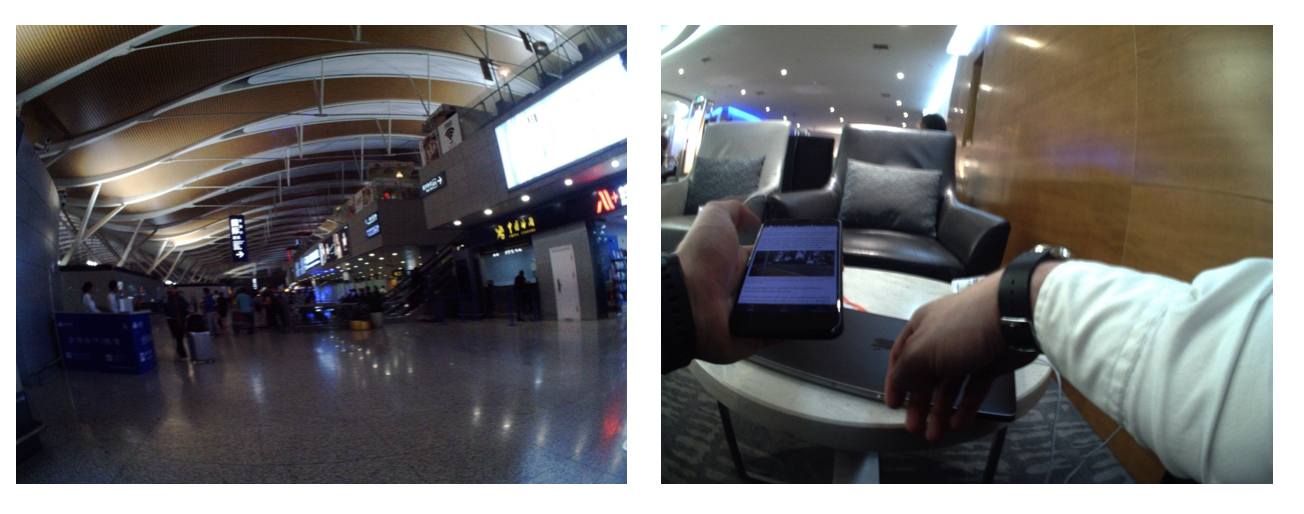}
  \caption{A sequence of first-person viewpoint images. }
  \label{fig:compare}
\end{figure}

Recent methods~\cite{10.1145/3592573.3593101,10.1145/3592573.3593105} address the semantic gap by using CLIP or other text-image embedding models to link images with text queries. 
However, CLIP models process images in isolation and don't consider temporal sequences, limiting their ability to accurately reconstruct a lifelogger’s experience. 
To bridge the gap further, some studies incorporate additional data, such as metadata or user queries. 
For example, \cite{10.1145/3512729.3533006} improve accuracy by ensembling cosine similarity scores, and \cite{10.1145/3512729.3533012} combine GPS, semantic locations, and temporal metadata with query parsing. 
\cite{10.1145/3592573.3593098} use query parsing for location and time to filter images, while \cite{cad7ae14c7f047749df68f0a7f74a70b} filter blurry images based on the variance of the Laplacian from grayscale.

We propose a Captioning-Integrated Visual Lifelog (CIVIL) Retrieval System using large video language models and large vision-language models (LVLMs) to tackle these challenges without model training.
Our approach bridges the semantic gap by converting images into textual captions. 
These captions, along with textual queries, are embedded into a unified vector space via a text embedding model.
Leveraging multi-modal language models' instruction-following abilities, the system adapts to lifelog images by framing prompts to reconstruct personal experiences. 
In this way, these models generate captions that reflect personal experiences in the visual lifelog. 
We propose three captioning methods, each utilizing different large multi-modal language models:
(1) the \textbf{Single Caption Method}, which generates captions for individual images; 
(2) the \textbf{Collective Caption Method}, which treats time-consecutive images as video frames and generates a unified caption; 
and (3) the \textbf{Merged Caption Method}, which produces fine-grained captions for each image and coarse-grained captions for groups of similar images, with the grouping determined by the LVLM.

    


These methods tackle the main challenges by using textual caption embeddings for image representation. 
We found that averaging similarity scores across different caption types improves retrieval performance with specific text embedding models. 
Our system outperforms direct image embedding methods, such as CLIP, as shown by experimental results. 
We conducted further analysis, including error analysis and an examination of the caption-query retrieval step facilitated by large language models (LLMs), to understand the reasons behind our system's success.
In sum, this paper contributes: 
\begin{enumerate}
    \item We present a simple system that represents lifelog images with captions, enabling retrieval through user queries.
    \item We provide a comprehensive analysis of our experimental results and explore ways to enhance our system further.
    \item We release the generated captions\footnote{https://github.com/ntunlplab/Visual-Lifelog-Retrieval-through-Captioning-Enhanced-Interpretation.git} on the NTCIR-14 Lifelog-3 dataset\cite{gurrin2019overview}, offering a valuable resource for future visual lifelog retrieval and understanding research.
\end{enumerate}

\section{Related Works}

Video Corpus Moment Retrieval (VCMR) involves retrieving specific segments from continuous video streams based on a text query, unlike visual lifelog retrieval, which uses periodically captured frames in lifelogging datasets. 
VCMR often utilizes unimodal encoding or cross-modal interaction learning.
\cite{Zhang_2021} presents ReLoCLNet, which trains video and text encoders via contrastive learning. 
While VCMR focuses on segment-based retrieval, visual lifelog retrieval emphasizes retrieving the top K images that capture query context.


Dense Video Captioning (DVC) aims to interpret events in videos and generate captions, involving video feature extraction, temporal event localization, and dense caption generation~\cite{qasim2023densevideocaptioningsurvey}.
Recent work~\cite{zhou2024streamingdensevideocaptioning} introduces a streaming model with a clustering memory module and streaming decoding for captioning long videos. 
Similarly, our system’s merged caption method uses GPT-4-turbo-vision~\cite{openai2024gpt4} to cluster lifelog images by event similarity and generate coarse-grained captions per group.
Unlike DVC, however, our system is designed for lifelog images, processing a fixed set of images at a time.


\section{Captioning Retrieval System}\label{sec:Methods}
\subsection{Overview}
The task is defined as follows: The system receives a set of lifelog videos, $(V_0, V_1, \ldots, V_N)$, where each video $V_i$ contains $M$ frames, $(F_i^0, F_i^1, \ldots, F_i^M)$.
The goal is to retrieve relevant frames from visual lifelogs based on a user query, selecting $K$ candidate frames to assist the user in recall.
Figure~\ref{fig:Sys_Arc} shows the system, which uses three caption generation methods: single, collective, and merged caption methods. 

\begin{figure*}[t]
  \centering
  \includegraphics[width=0.95\textwidth]{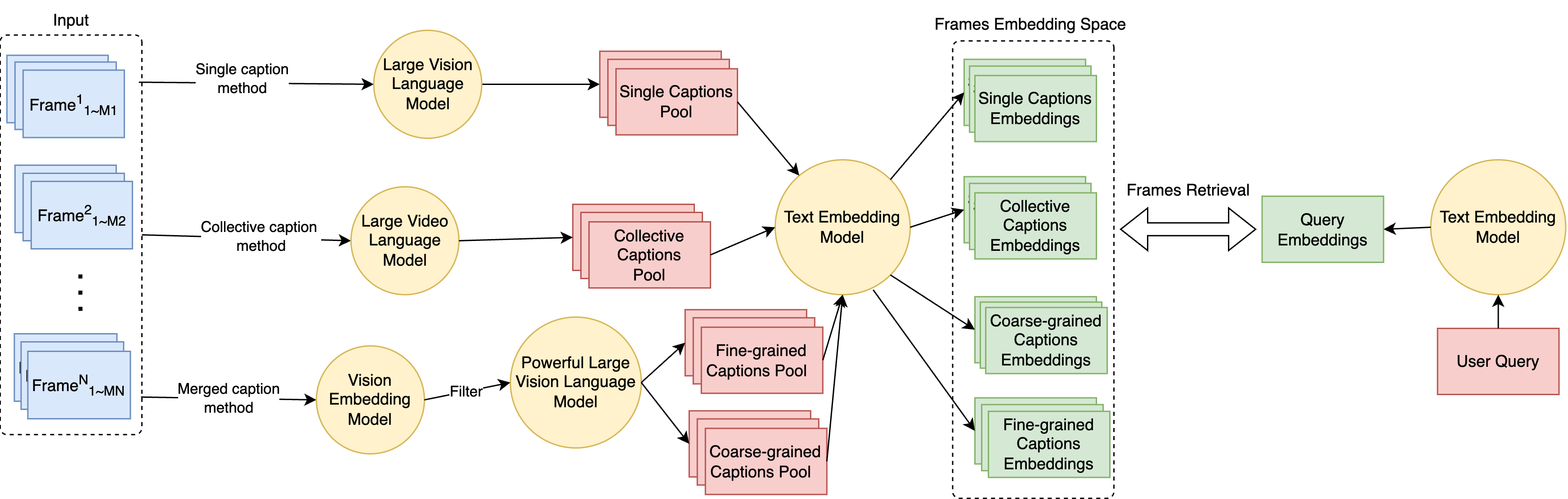}
  \caption{CIVIL Retrieval System Architecture. The system processes frames from multiple videos using three methods: (1) single-frame captions by an LVLM, (2) collective captions by a large video language model, and (3) fine- and coarse-grained captions by a Vision Embedding Model and LVLM. All captions and the user query are embedded in a shared text embedding space, and relevant frames are retrieved based on query-caption similarity.}
  \label{fig:Sys_Arc}
\end{figure*}

\subsection{Single caption method}\label{sec:Single}
The single caption method generates a caption for each frame using an LVLM, guided by a prompt that includes the frame and metadata. 
The prompt aims to reconstruct personal experiences with captions limited to 20-40 words, highlighting key moments, locations, and activities. 
Due to the low quality of images and the lack of context from neighboring frames, we assume that single-frame captions may be prone to hallucinations and are time-consuming to produce. 
To this end, we propose the collective caption method.

\subsection{Collective caption method}\label{sec:Collec}
The collective caption method uses a large video language model to generate a caption for a set of consecutive frames, incorporating metadata (e.g., time interval) and extending the caption length to 40-60 words for comprehensiveness.
Collective captions offer broader descriptions than single captions but may lack detail. 
Retrieving a collective caption brings in all associated images, make it difficult to identify the relevant one in the sequence.
Our combination experiment (see Section~\ref{sec:combination}) provides a solution.

\subsection{Merged caption method}\label{sec:Merged}
To leverage both single and collective caption strengths, we use GPT-4-turbo-vision to handle complex prompts. 
Due to its high cost, we first apply a filtering algorithm that embeds each frame, calculates similarity scores, and retains frames where consecutive pairs exceed a set similarity threshold.



The input comprises a fixed number of filtered frames, metadata (time info), and a complex instruction prompt with three tasks: 
    (1) \textbf{Fine-grained caption:} Provide brief captions for each image with metadata. 
    (2) \textbf{Frame summary:} Summarize the individual's experiences in 20-40 words. 
    (3) \textbf{Coarse-grained caption:} Group frames and describe each group. 

To ensure experience consistency among model's outputs, we include the frame summary from the previous output in the current prompt. 
We assume that fine-grained captions provide detailed descriptions for each frame while minimizing hallucinations with additional contextual information. 
However, the complexity of this prompt design requires a powerful LVLM, which may require additional computational resources.

\subsection{Frame Retrieval From User Query}\label{sec:Retrieval}

 We employ a text embedding model to project generated captions and user queries into a shared vector space. Four types of captions—single, collective, fine-grained, and coarse-grained—are embedded. To enhance matching with user queries, we prepend "The individual’s experience:" before embedding the captions. User queries are embedded similarly, and similarity scores between queries and captions are computed. Based on these scores, we retrieve the top $K$ captions most similar to the query. Single and fine-grained captions correspond to individual frames (filtered frames for fine-grained captions), while collective and coarse-grained captions correspond to multiple frames (multiple filtered frames for coarse-grained captions).

 Combining the similarity scores of single or fine-grained captions with those of collective or coarse-grained captions improves performance. 
Section~\ref{sec:combination} details the methods. 

\section{Experiments Setup}\label{sec:experiment}






\subsection{Dataset and Evaluation Metrics}\label{sec:dataset}
In our experiment, we use images from the NTCIR-14 Lifelog-3 Dataset~\cite{gurrin2019overview}, focusing on data from 'User 1,' which includes approximately 63,000 images captured over 29 days. 
We primarily use timestamps, assuming they are relevant for identifying activities such as eating, waking, or sleeping.

To evaluate retrieval performance, we use the topics, labels, and evaluation metrics from ImageCLEF 2019 LMRT~\cite{ImageCLEF19}, which involves retrieving relevant images based on 10 user-query topics. 
The task emphasizes diversity in the retrieved images to avoid selecting similar ones from short consecutive intervals. The queries, provided by ImageCLEF, are designed to identify specific activities or moments from the lifelog data.


The evaluation metrics used in this task—Cluster Recall at $K$ ($CR@K$), Precision at $K$ ($P@K$), and F1 at $K$ ($F1@K$)—are based on the definitions established by the ImageCLEF 2019 LMRT competition organizers. 
Our primary metric is average $P@10$, with additional reports on $CR@10$ and $F1@10$ for further performance analysis. 





\subsection{Baseline Models}\label{sec:baseline_model}
We compare our CIVIL Retrieval System with standard text-based image retrieval methods, which directly embed images and user queries into a shared vector space and retrieve images based on similarity scores, without using caption transformation.
The baseline models used for comparison are: CLIP ViT-B/32, CLIP ViT-L/14~\cite{radford2021learning}, {ViT-SO400M-14-SigLIP-384~\cite{zhai2023sigmoid}, and DFN5B-CLIP-ViT-H-14~\cite{fang2023data}.




\subsection{Caption Models}\label{sec:caption_model}
For the methods in Sections~\ref{sec:Single}, \ref{sec:Collec}, and \ref{sec:Merged}, we select different language and vision models.

\subsubsection{Single Caption Model}\label{single_model}
\textcolor{black}{We implement the Caption Retrieval Method using several LVLMs.} 
After generating captions, we use a text embedding model to rank and retrieve images based on similarity scores between captions and user queries.
The models used for this task include Idefics2~\cite{laurençon2024matters}, LLaVA-NeXT-7B~\cite{liu2024llavanext}, and Internlm-xcomposer2-vl-7B~\cite{dong2024internlmxcomposer2}.


    

\subsubsection{Collective Caption Model}\label{sec:collec_model}
In this approach, lifelog images are treated as video frames, and a collective caption is generated for every set of eight consecutive frames. Images are then retrieved based on similarity scores between these collective captions and user queries. Retrieving a single collective caption corresponds to retrieving a set of eight images, ensuring that if the caption accurately describes the content, all relevant images in the set are included.

The large video language models used include Video-LLaVA~\cite{lin2023videollava} and Large-World-Model (LWM)~\cite{liu2024world}.



\subsubsection{Merged Caption Model}
We employ GPT-4-turbo-vision\footnote{Version released on 2023-11-06} to generate fine-grained and coarse-grained captions and summaries of lifelog images in JSON format. Ten consecutive filtered images are provided to the model. The fine-grained and coarse-grained captions are used for retrieval, similarly to the processes for single and collective captions, respectively. The key difference is that a retrieved fine-grained caption is linked to a specific filtered image, whereas a coarse-grained caption corresponds to all filtered images in the group.
\subsection{Vision Embedding Model and Text Embedding Model}\label{sec:auxiliary_model}

\noindent \textbf{Vision Embedding Model:} 
We use the DINOv2 ViT-G14 model~\cite{oquab2024dinov2} for filtering algorithm in merged caption Method. We compute the cosine similarity between embedded images and set a threshold of 0.8.

\noindent \textbf{Text Embedding Model:} 
To rank and retrieve images based on their generated captions and user queries, we use two text embedding models:
GTE-Large\cite{li2023general} and BGE-M3\cite{chen2024bge}.



\subsection{Combination Captions Experiment}\label{sec:combination}
We investigate whether combining fine-grained and coarse-grained captions, as well as single and collective captions, can enhance retrieval performance by providing more comprehensive information. Four combinations of similarity scores are tested using two different text embedding models:
(1)~combining fine-grained and coarse-grained captions with each embedding model, and
(2)~combining single and collective captions generated by the models with the highest average P@10 scores with each embedding model.
The combination similarity score is calculated by averaging the similarity scores of the components.


\section{Experimental Results}


\subsection{Overall Performance Results}
Table~\ref{tab:gte_metrics} presents baseline and caption model performance with GTE-large and BGE-M3 text embedding models.

\subsubsection{Precision at 10 (P@10) Results}

Among baseline models, ViT-SO400M-14-SigLIP-384 and DFN5B-CLIP-ViT-H-14 achieved the highest P@10, averaging 0.58. Notably, five of nine caption methods surpassed this baseline with GTE-large embeddings, while only three did so with BGE-M3 embeddings. The highest P@10 of 0.73 was achieved by combining InternLM-XComposer2-VL-7B and Video-LLaVA under GTE-large embeddings, indicating that 73 out of 100 retrievals were accurate from a pool of 63,000 images.

The performance across different caption methods showed considerable variation.
The single caption methods generally performed well, with LLaVA-NeXT-7B using BGE-M3 achieving the top P@10 score of 0.71. 
In contrast, collective caption methods showed mixed results: Video-LLaVA slightly underperformed the baseline with GTE-large embeddings, and LWM performed poorly across both embeddings. 
The merged caption method, despite using GPT-4-turbo-vision, did not exceed the baseline performance for either text embedding model, with potential performance-reducing factors discussed in Section~\ref{sec: merged_analysis}. 
Finally, the combination method, which merged collective and single caption scores, achieved the highest performance of P@10 at 0.73 with GTE-large embeddings. 
However, combining merged caption methods under BGE-M3 embeddings reduced the performance significantly, dropping to a P@10 of 0.47 from 0.57 in coarse-grained caption retrieval.

In summary, averaging similarity scores of BGE-M3 embeddings from two captions did not outperform using GTE-large embeddings. Subsequent analysis focuses on the Combination method with the GTE-large model.


\subsubsection{Cluster Recall at 10 (CR@10) and F1 at 10 (F1@10) Results}

Without employing diversity-promoting algorithms, our system shows potential in CR@10 and F1@10 metrics. The LLaVA-NeXT-7B model achieves a CR@10 of 0.572, indicating that 57.2\% of the correct clusters are included in the top 10 retrievals per topic. With a P@10 of 0.71, it attains an F1@10 of 0.541. Further investigation is needed to understand the factors contributing to this performance.

\subsection{Topic-Specific Results}
P@10 scores across individual topics reveal varying degrees of difficulty and model effectiveness. These observations are specific to our dataset and should not be generalized.

In Topic~4, with only 3 correct images, baseline models performed relatively better, typically retrieving 1 correct image out of 10, while most other methods, except the single caption method, failed to retrieve any correct images. Conversely, in Topic~1, with 16 correct images, some methods showed notable success. For instance, the GPT-4 coarse-grained caption method with BGE-M3 embeddings retrieved all 10 correct images. In Topic 8, with 40 correct images, model effectiveness varied. The Video-LLaVA model combined with GTE-large embeddings retrieved all 10 correct images, and 13 out of 22 models retrieved more than 7 correct images across both embeddings and baseline models.
In topics with fewer than 1,000 correct images (Topics 5, 6, 7, 9), performance varied. Topics 5 and 6 were relatively easier, with most models retrieving 8 to 10 correct images. Topic 9 was more challenging; baseline models, collective caption methods, and merged caption methods retrieved no more than 4 correct images. In Topic 7, most methods retrieved over 6 correct images.
In topics with more than 1,000 correct images (Topics 2, 3, 10), performance was mixed. Topic 3 posed significant challenges, with many models retrieving no more than 4 correct images. For the other topics, more than half of the models retrieved over 6 correct images.


Difficulty in certain topics might be linked to blurry images in the lifelog dataset. For example, in Topic 3, blurry images could hinder distinguishing screens displaying videos from those engaged in other tasks, thereby reducing P@10 scores.

\begin{table*}[t]
\centering
\begin{tabular}{l|c|c|c|c|c|c|c|c|c|c|c|c|c}
\hline
Method Name & T1 &T2 & T3 & T4 & T5 & T6 & T7 & T8 & T9 & T10 & Avg CR@10 & Avg F1@10 & Avg P@10 \\
\hline
\multicolumn{14}{l}{Baseline} \\
\hline
CLIP ViT-B/32 & 0.4 & \textbf{1.0} & 0.1 & 0.1 & 0.9 & 0.9 & 0.7 & 0.9 & 0.3 & 0.4 & 0.522 & 0.447 & 0.57 \\
CLIP ViT-L/14 & 0.5 & 0.8 & 0.7 & 0.1 & \textbf{1.0} & 0.8 & 0.3 & 0.4 & 0.0 & 0.8 & 0.475 & 0.37 & 0.54 \\
ViT-SO400M-14-SigLIP-384 & 0.7 & \textbf{1.0} & 0.0 & 0.1 & \textbf{1.0} & 0.6 & 0.7 & 0.7 & 0.4 & 0.6 & 0.506 & 0.448 & 0.58 \\
DFN5B-CLIP-ViT-H-14 & 0.4 & \textbf{1.0} & 0.0 & 0.1 & \textbf{1.0} & 0.7 & 0.7 & 0.7 & 0.4 & 0.8 & 0.460 & 0.407 & 0.58 \\
\hline
\multicolumn{14}{l}{Single caption method with GTE-large text embedding} \\
\hline
Idefics2 & 0.7 & 0.7 & 0.1 & 0.0 & \textbf{1.0} & 0.8 & 0.7 & 0.8 & 0.7 & 0.5 & 0.446 & 0.478 & 0.6 \\
LLaVA-NeXT & 0.5 & 0.9 & 0.4 & \textbf{0.2} & \textbf{1.0} & \textbf{1.0} & 0.6 & 0.7 & 0.0 & 0.6 & 0.518 & 0.460 & 0.59 \\
Internlm-xcomposer2-vl-7B & 0.5 & 0.9 & 0.4 & 0.1 & \textbf{1.0} & 0.8 & 0.6 & 0.9 & 0.5 & \textbf{0.9} & 0.506 & 0.484 & 0.66 \\
\hline
\multicolumn{14}{l}{Collective caption method with GTE-large text embedding} \\
\hline
Video-LLaVA & 0.4 & \textbf{1.0} & 0.0 & 0.0 & \textbf{1.0} & 0.8 & \textbf{0.9} & \textbf{1.0} & 0.0 & 0.6 & 0.245 & 0.278 & 0.57 \\
LWM & 0.0 & \textbf{1.0} & 0.0 & 0.0 & \textbf{1.0} & 0.0 & 0.9 & 0.0 & 0.0 & 0.0 & 0.060 & 0.099 & 0.29 \\
\hline
\multicolumn{14}{l}{Merged caption method with GTE-large text embedding} \\
\hline
GPT4\_fine-grained & 0.7 & 0.9 & 0.5 & 0.0 & 0.9 & 0.8 & 0.6 & 0.6 & 0.4 & 0.4 & 0.482 & 0.485 & 0.58 \\
GPT4\_coarse-grained & 0.8 & 0.9 & 0.7 & 0.0 & 0.9 & \textbf{1.0} & 0.2 & 0.4 & 0.1 & 0.3 & 0.292 & 0.321 & 0.53 \\
\hline
\multicolumn{14}{l}{GTE-large text embedding Combination Method} \\
\hline
Internlm $\times$ Video-LLaVA & 0.5 & 0.9 & 0.2 & 0.0 & \textbf{1.0} & \textbf{1.0} & \textbf{0.9} & 0.9 & \textbf{1.0} & \textbf{0.9} & 0.406 & 0.478 & \textbf{0.73} \\
fine-grained $\times$ coarse-grained & 0.9 & \textbf{1.0} & 0.6 & 0.0 & \textbf{1.0} & \textbf{1.0} & 0.5 & 0.3 & 0.5 & 0.5 & 0.426 & 0.475 & 0.63 \\
\hline
\multicolumn{14}{l}{Single caption method with BGE-M3 text embedding} \\
\hline
Idefics2 & 0.3 & 0.7 & 0.2 & 0.1 & \textbf{1.0} & \textbf{1.0} & 0.6 & 0.7 & 0.7 & 0.3 & 0.532 & 0.440 & 0.56 \\
LLaVA-NeXT & 0.8 & 0.9 & 0.5 & 0.1 & \textbf{1.0} & 0.9 & \textbf{0.9} & 0.8 & 0.3 & \textbf{0.9} & \textbf{0.572} & \textbf{0.541} & 0.71 \\
Internlm-xcomposer2-vl-7B & 0.6 & \textbf{1.0} & 0.3 & 0.0 & 0.9 & 0.9 & 0.5 & 0.8 & 0.7 & \textbf{0.9} & 0.476 & 0.515 & 0.66 \\
\hline
\multicolumn{14}{l}{Collective caption method with BGE-M3 text embedding} \\
\hline
Video-LLaVA & 0.4 & 0.2 & 0.0 & 0.0 & \textbf{1.0} & 0.8 & 0.3 & 0.8 & 0.3 & 0.8 & 0.332 & 0.316 & 0.46 \\
LWM & 0.0 & \textbf{1.0} & 0.8 & 0.0 & \textbf{1.0} & 0.0 & 0.6 & 0.0 & 0.2 & 0.4 & 0.084 & 0.127 & 0.40 \\
\hline
\multicolumn{14}{l}{Merged caption method with BGE-M3 text embedding} \\
\hline
GPT4\_fine-grained & 0.7 & 0.8 & 0.4 & 0.0 & 0.6 & 0.9 & 0.4 & 0.5 & 0.1 & 0.7 & 0.422 & 0.422 & 0.51 \\
GPT4\_coarse-grained & \textbf{1.0} & 0.8 & \textbf{0.9} & 0.0 & \textbf{1.0} & \textbf{1.0} & 0.7 & 0.0 & 0.0 & 0.3 & 0.241 & 0.299 & 0.57 \\
\hline
\multicolumn{14}{l}{BGE-M3 text embedding Combination Method} \\
\hline
LLaVA-NeXT $\times$ Video-LLaVA & 0.7 & 1.0 & 0.3 & 0.0 & \textbf{1.0} & \textbf{1.0} & 0.6 & 0.9 & 0.7 & 0.8 & 0.442 & 0.497 & 0.7 \\
fine-grained $\times$ coarse-grained & 0.9 & 0.7 & 0.1 & 0.0 & 0.9 & \textbf{1.0} & 0.3 & 0.4 & 0.1 & 0.3 & 0.426 & 0.406 & 0.47 \\
\hline
\end{tabular}
\caption{Results of baseline and our CIVIL Retrieval System. Scores for Topic1 (T1) to Topic10 (T10) represent P@10.}
\label{tab:gte_metrics}
\end{table*}

\section{Discussion}\label{sec:discuss}

\subsection{Error Analysis}
We document errors that result in both the exclusion of correct images from the top 10 and the inclusion of incorrect ones across 10 topics. For Topic 4, with only three correct images, we focus on the top 3 retrieved images. 
For the CIVIL Retrieval System, we use the text embedding variant with the highest P@10 score, excluding the combination method (see Section~\ref{sec:comb_analysis}), to reduce text embedding impact on results.
BGE-M3 text embedding is used for the Internlm-xcomposer2-vl-7B model. 
Our analysis is limited to the top 10 retrievals for each topic, ignoring errors in lower-ranked results.



\subsubsection{Error Analysis of Image Embedding Retrieval Baseline}\label{sec:subsubimage}

We analyze retrieval errors on ViT-SO400M-14-SigLIP-384 and DFN5B-CLIP-ViT-H-14 for their superior performance. As the embedding models' mechanisms are not explicitly known, the following error types are based on our assumptions.

\noindent \textbf{Contextual Image Error}: This error arises due to the model's limitation of processing a single image without considering the context from neighboring frames. For instance, in Figure~\ref{fig:contextual_images} (a)(b)(c), recognizing the action in image (c) (such as eating ice cream) would require contextual information from images (a) and (b). Similarly, in sequence (d)(e)(f), determining whether the person in (e) is reading a book or retrieving it from a printer depends on the context provided by (d) and (f).


\noindent \textbf{First-Person Viewpoint Error}: This error occurs when actions performed by individuals in the background are mistakenly attributed to the lifelogger. For instance, in Figure~\ref{fig:caption_image_err}(a), a person taking a photo on a bridge may be misidentified as the lifelogger, leading to an incorrect association with Topic~4.


\noindent \textbf{Object Detection Hallucination}: Error arising from object misidentification. As shown in Figure~\ref{fig:caption_image_err}(b), an image is retrieved for Topic~1 possibly because a food item is incorrectly identified as ice cream, resulting in an erroneous retrieval.

\noindent \textbf{Event Detection Error}: Error arising from event action misidentification. For example, Figure~\ref{fig:caption_image_err} (c) is retrieved for Topic 7, even though the depicted activity (eating food) does not match the topic description (cooking).

\noindent \textbf{Labeling Error}: Error stemming from incorrect labeling in the dataset. In Figure~\ref{fig:caption_image_err} (d), the image is not labeled under Topic 9, although it shows a person in a shuttle bus at an airport, which matches the topic's description.

\noindent \textbf{Interpretative Labeling Error}: This error arises due to subjective differences in the interpretation of the dataset's labels. For instance, Figure~\ref{fig:caption_image_err} (e) is not labeled under Topic 8, but it could be considered as “in” a car salesroom based on varying criteria. Such errors differ from standard labeling errors as they depend on individual interpretation.

Table~\ref{tab:cap_analysis} presents the frequency of each error type across both models for the top 10 retrievals over 10 topics. Event detection errors are the most frequent, particularly impacting the accuracy of image-query text pairing. Contextual image errors are also notable, especially in the DFN5B-CLIP-ViT-H-14 model. For a fair comparison of performance between baseline models and our system, we provide a corrected evaluation on average P@10 in Table~\ref{tab:correct_metrics}.

    }

\begin{figure}[htbp]
    \centering
    \begin{subfigure}[b]{0.13\textwidth}
        \includegraphics[width=\textwidth]{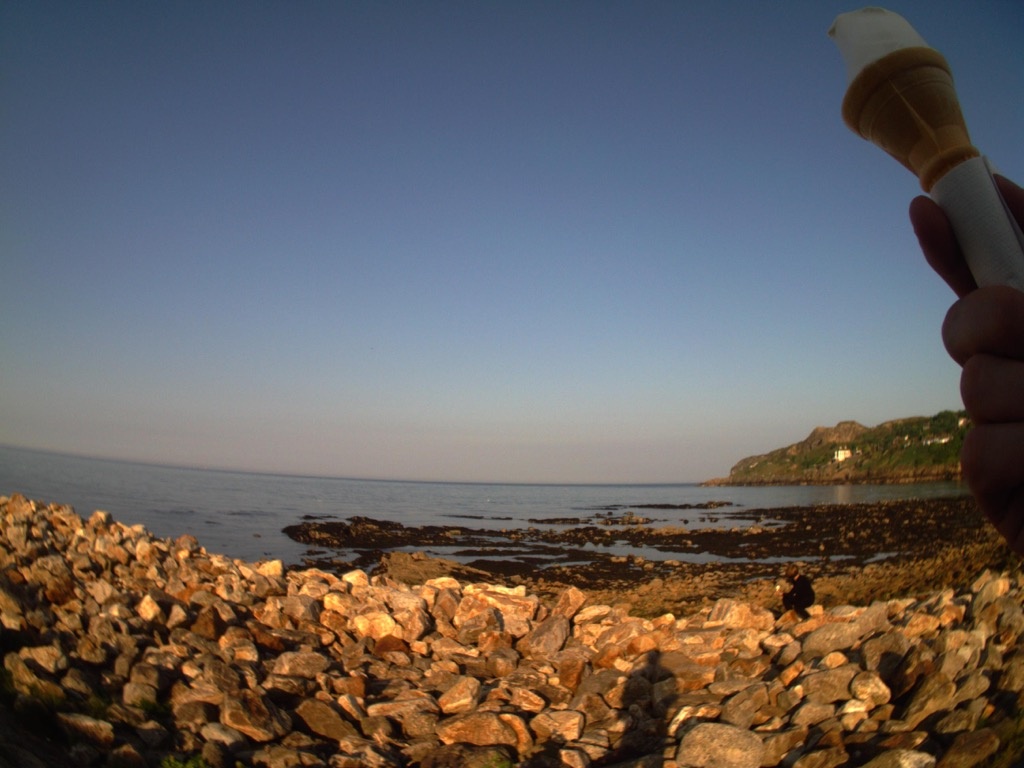}
        \caption{}
    \end{subfigure}
    \begin{subfigure}[b]{0.13\textwidth}
        \includegraphics[width=\textwidth]{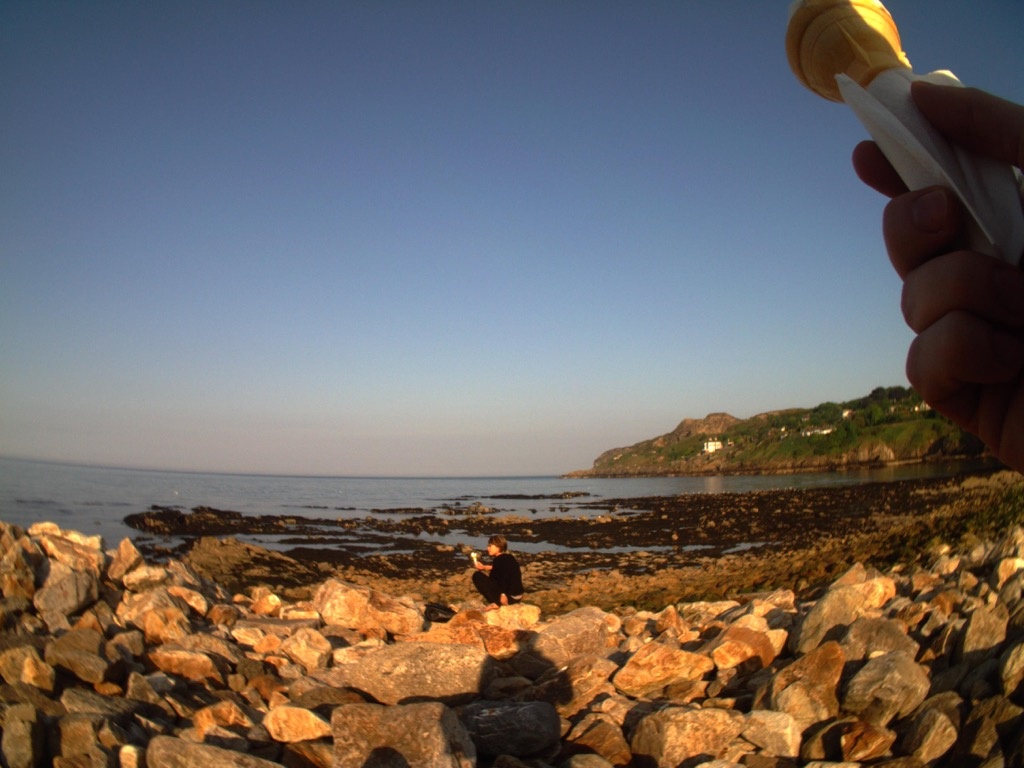}
        \caption{}
    \end{subfigure}
    \begin{subfigure}[b]{0.13\textwidth}
        \includegraphics[width=\textwidth]{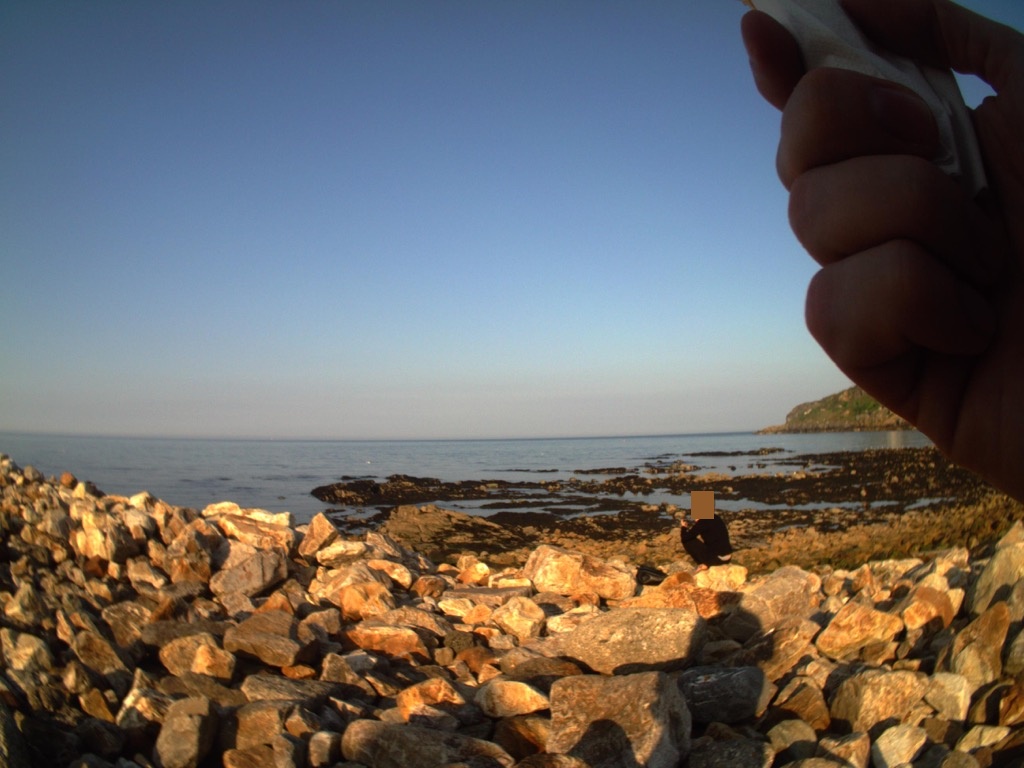}
        \caption{}
    \end{subfigure}
    
    \begin{subfigure}[b]{0.13\textwidth}
        \includegraphics[width=\textwidth]{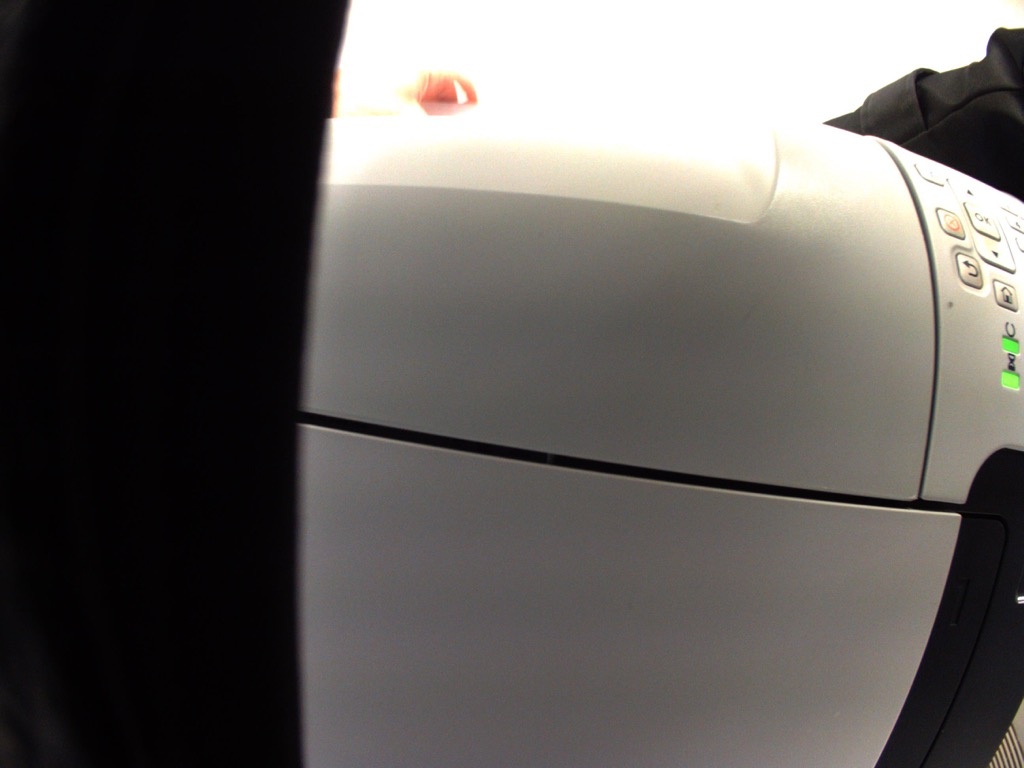}
        \caption{}
    \end{subfigure}
    \begin{subfigure}[b]{0.13\textwidth}
        \includegraphics[width=\textwidth]{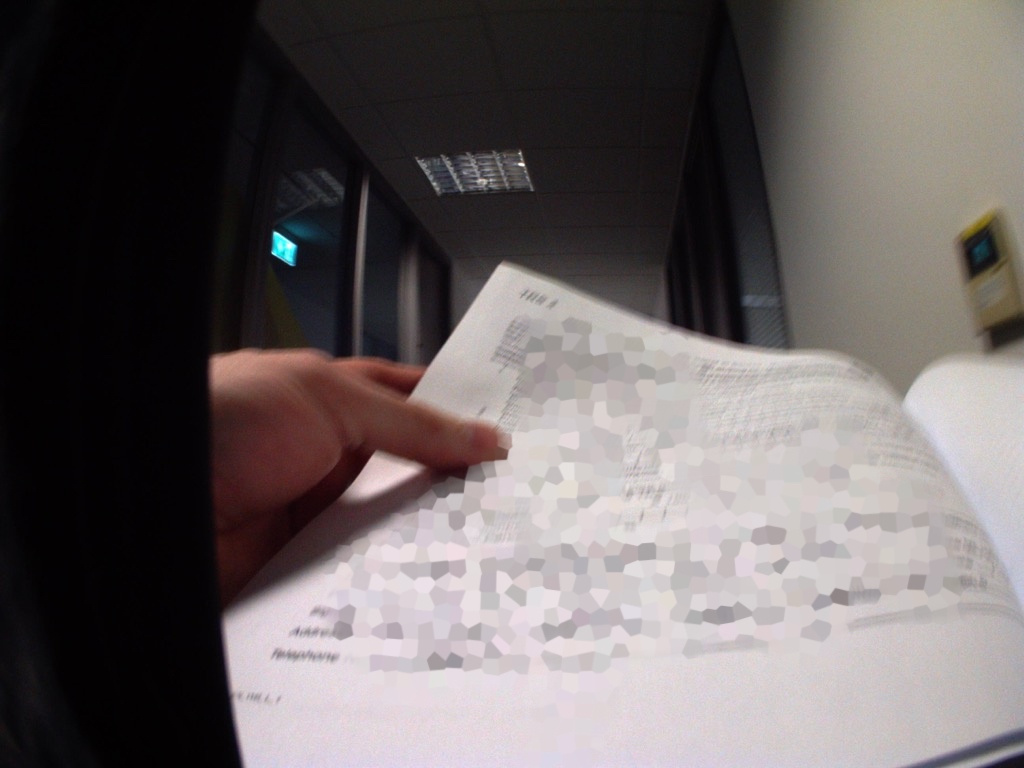}
        \caption{}
    \end{subfigure}
    \begin{subfigure}[b]{0.13\textwidth}
        \includegraphics[width=\textwidth]{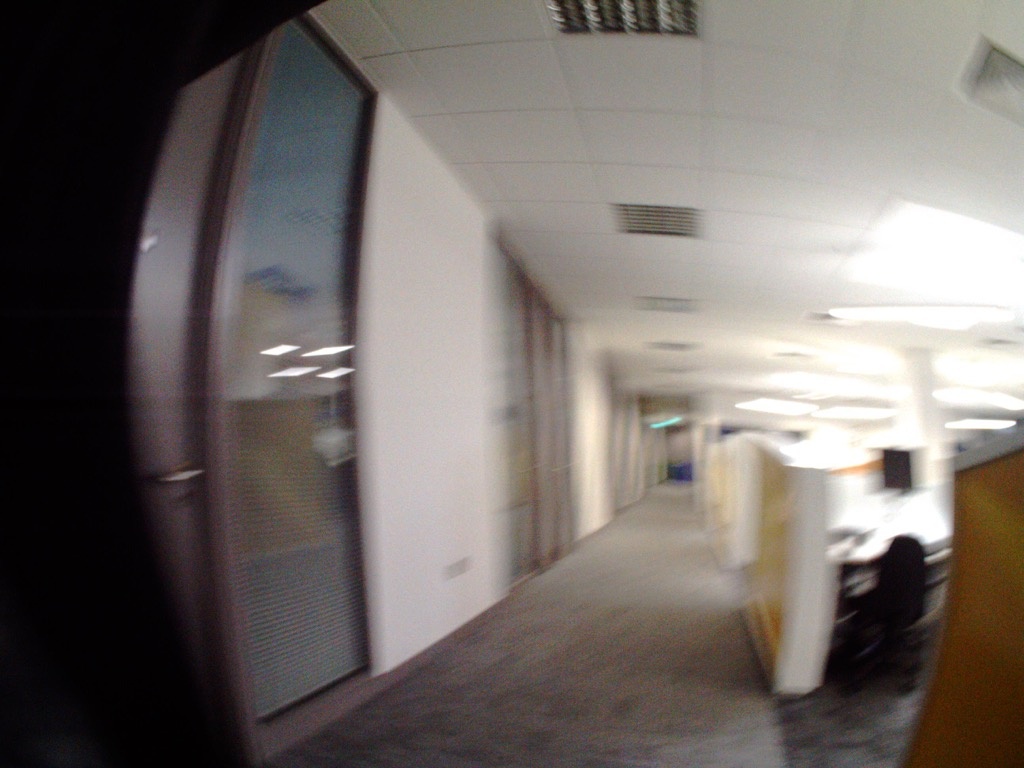}
        \caption{}
    \end{subfigure}
    
    \caption{Examples of Contextual Images Problem}
    \label{fig:contextual_images}
\end{figure}

\subsubsection{Analysis of CIVIL Retrieval Methods}

CIVIL retrieval methods convert images into textual captions before retrieval, allowing for a more interpretable analysis. While they share errors similar to those observed in the baseline model—such as contextual image errors, first-person viewpoint errors, event detection errors, object detection hallucinations, labeling errors, and interpretative labeling errors—they also present unique challenges specific to the captioning approach. The following errors are specific to caption retrieval methods:



\noindent \textbf{Text Embedding Retrieval Error}: Error in which the caption of an incorrect image ranks higher in retrieval due to poor text embedding performance or lack of better caption candidates in the pool. For instance, the BGE-M3 text embedding model ranks the sentence ``Driving a car, possibly starting a journey'' closer to Topic 9 than a more accurate description ``The series of images show the individual engaging with their smartphone and holding a coffee cup while traveling on a bus''.


\noindent \textbf{Lack of Specific Details}: This error occurs when a caption omits essential details, such as location or activity information, that are required by the prompt. For instance, Idefics2 generates a vague caption, ``In this image I can see a screen and a device,'' for Figure~\ref{fig:caption_image_err}(f), without specifying that it depicts an airplane cabin. Note that detecting this error can be subjective.
    
\noindent \textbf{Over-Interpretation Error}: This error occurs when caption models infer or associate details that are not clearly visible in the image. For example, LLaVA-NeXT-7B generates the caption for Figure~\ref{fig:caption_image_err}(g): ``… as a person is about to enjoy a bag of chips. The individual, whose hand is visible in the foreground, is holding the bag with a sense of eagerness. The bag is from Fresh Fish \& Chips, \textit{suggesting that the person might be at a seaside location}…'' This caption speculates about the setting, incorrectly associating a bag of chips with a “seaside location,” which is an over-interpretation.


\noindent \textbf{Error Propagation}: Specific to collective caption and coarse-grained caption retrieval, this error arises when each retrieved caption represents multiple images, and errors in that caption propagate across all associated images. Unlike single-image captioning, this can lead to repeated errors in retrieval.

\noindent \textbf{Fixed Frame Limitation}: This error occurs when a model processes a fixed number of frames (e.g., 8). For example, if only \textit{image\_1} to \textit{image\_3} are correct, the 8-frame window would includes 5 irrelevant frames. This limitation affects coarse-grained and collective caption retrieval.

Table~\ref{tab:cap_analysis} summarizes the error occurrences for different models. 
Text-embedding retrieval and event detection errors are the main bottlenecks, relying heavily on text embedding accuracy and precise image captions. 
Unlike the collective caption method, coarse-grained retrieval reduces error propagation by dynamically grouping images and avoiding irrelevant additions.
Table~\ref{tab:correct_metrics} presents the P@10 results after removing label errors and interpretative labeling errors. Notably, 4 out of 10 captioning methods perform better than the baseline models, with LLaVA-NeXT-7B and its combination with Video-LLaVA achieving the highest score of 0.78.

\subsubsection{Analysis of Merged Caption Method}\label{sec: merged_analysis}
GPT-4-turbo-vision effectively generates fine-grained captions by using context from neighboring images. 
However, relying on previous summaries for subsequent captions can cause error propagation if inaccuracies occur. 
Interestingly, coarse-grained captions and experience summaries excel in reasoning across multiple images, suggesting that concatenating these summaries could reconstruct personal daily experiences.

\begin{table*}[t]
    \centering
    \small
    \setlength{\tabcolsep}{3pt}
    \begin{tabularx}{\textwidth}{l*{11}{>{\centering\arraybackslash}X}}
        \hline
        & (a) & (b) & (c) & (d) & (e) & (f) & (g) & (h) & (i) & (j) & (k) \\
        \hline
        ViT-SO400M-14-SigLIP-384 & 4 & 2 & - & 18 & - & - & 3 & 10 & 3 & - & - \\
        DFN5B-CLIP-ViT-H-14 & 8 & 1 & - & 19 & - & - & 4 & 7 & 3 & - & -\\
        \hline
        
        Single Caption Method: Idefics2 & 8 & 2 & 11 & 14 & 2 & 0 & 0 & 5 & 4 & - & - \\
        Single Caption Method: LLaVA-NeXT & 3 & 0 & 4 & 6 & 0 & 3 & 3 & 5 & 2 & - & - \\
        Single Caption Method: Internlm-xcomposer2-vl-7B & 8 & 0 & 13 & 11 & 1 & 1 & 0 & 1 & 2 & - & - \\
        \hline
        
        Collective Caption Method: Video-LLaVA & - & 0 & 4 & 2 & 3 & 0 & 0 & 0 & 0 & 39 & 5 \\
        Collective Caption Method: LWM & - & 0 & 7 & 10 & 2 & 0 & 0 & 1 & 0 & 64 & 4 \\
        \hline
        
        Merged Caption Method: GPT4\_fine-grained & - & 1 & 15 & 11 & 9 & 0 & 1 & 8 & 6 & - & - \\
        Merged Caption Method: GPT4\_coarse-grained & - & 0 & 10 & 8 & 1 & 1 & 0 & 3 & 6 & 14 & 2 \\
        \hline
    \end{tabularx}
    \caption{Error occurrences among different methods. (a) Contextual Images Error (b) First-Person Viewpoint Error (c) Text Embedding Retrieval Problem (d) Event Detection Error (e) Lack of Specific Details (f) Over-Interpretation Error (g) Object Detection Hallucination (h) Labeling Error (i) Interpretative Labeling Error (j) Errors Propagation (k) Fixed Frame Limitation}
    \label{tab:cap_analysis}
\end{table*}

\begin{figure}[t]
  \centering
  \includegraphics[width=1\linewidth]{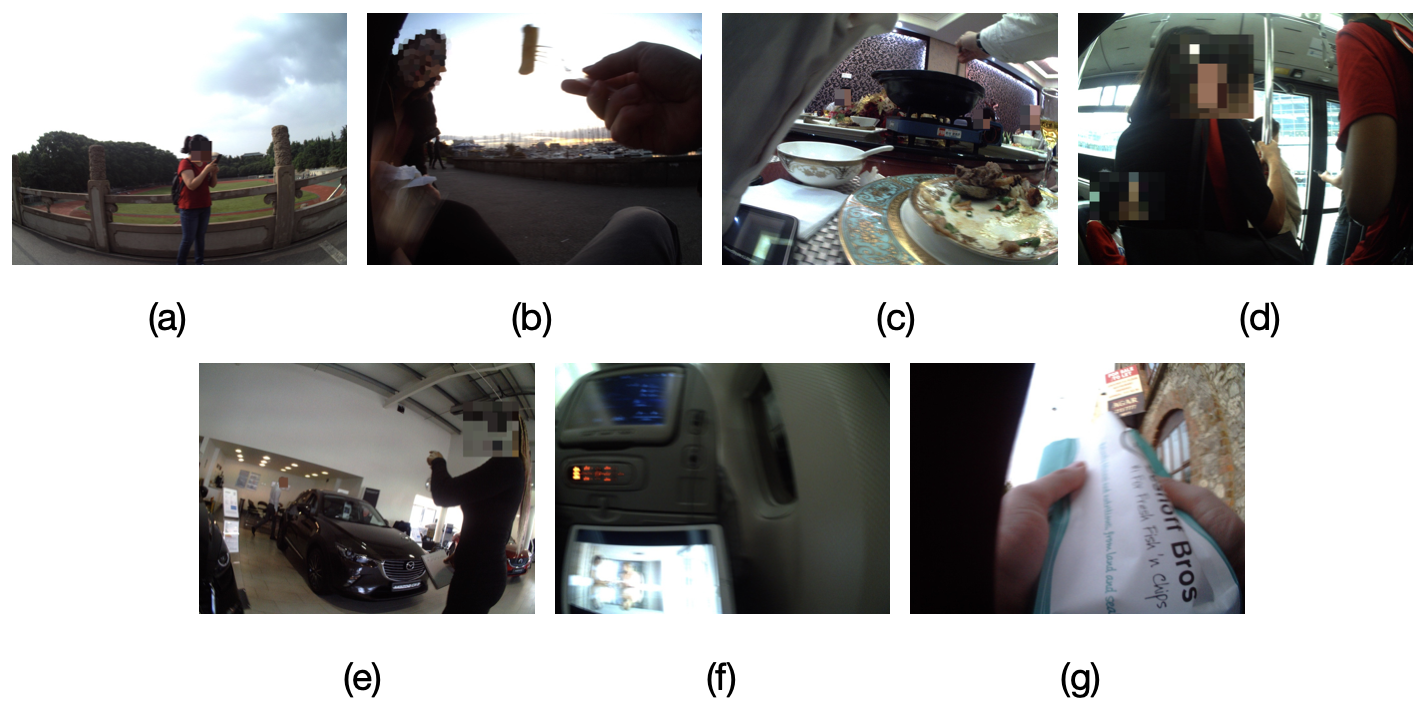}
  \caption{Examples of errors across different methods.}
  \label{fig:caption_image_err}
\end{figure}

\begin{table}[t]
\centering
\small 
\begin{tabular}{p{4.4cm} @{\hskip 4pt} c @{\hskip 4pt} c} 
\hline
Method & \makecell{Corrected \\ Avg P@10} & \makecell{Text Embedding \\ Model} \\
\hline

ViT-SO400M-14-SigLIP-384 & 0.71 & - \\
DFN5B-CLIP-ViT-H-14 & 0.68 & - \\
\hline
Idefics2 & 0.69 & GTE-large \\
LLaVA-NeXT & 0.78 & BGE-M3 \\
Internlm-xcomposer2-vl-7B & 0.69 & BGE-M3 \\
\hline
Video-LLaVA & 0.57 & GTE-large \\
LWM & 0.41 & BGE-M3 \\
\hline
GPT4\_fine-grained & 0.72 & GTE-large \\
GPT4\_coarse-grained & 0.66 & BGE-M3 \\
\hline
Internlm $\times$ Video-LLaVA & 0.76 & GTE-large \\
LLaVA-NeXT $\times$ Video-LLaVA & 0.78 & BGE-M3 \\
fine-grained $\times$ coarse-grained & 0.70 & GTE-large \\
\hline
\end{tabular}
\caption{Correction of the label error and the problem of error depending on individual interpretation.}
\label{tab:correct_metrics}
\end{table}

\subsection{Improvement of Text Retrieval Process}\label{sec:text_improvement}


We conducted an experiment using GPT-4o to identify the 10 most similar captions from the top 100 initially retrieved using the GTE-large text embedding model, selected for its superior performance with fine-grained captions. GPT-4o's fine-grained approach was chosen due to a higher incidence of retrieval issues with text embeddings. To leverage GPT-4o's strong instruction-following capabilities, dataset-provided topic descriptions were included in the prompt.

The result of incorporating GPT-4o shows notable improvements. The P@10 scores for Topic 1 through Topic 10 were 0.7, 0.9, 0.5, 0.0, 1.0, 1.0, 0.6, 0.5, 0.9, and 0.5, respectively. 
This improved the average P@10 from 0.58 to 0.66, and the corrected P@10 from 0.72 to 0.79, highlighting the effectiveness of GPT-4o in aligning captions with queries.
However, due to the high cost of using GPT-4o, future enhancements may involve exploring combination of text embedding model and reranker model to improve retrieval accuracy.


\subsection{Impact of Combination Method}\label{sec:comb_analysis}
We analyze the replacement effects from combining similarity scores using GTE-large embeddings, focusing on cases where an image with the lowest similarity score—whether from a single or fine-grained caption—enters the top 10 retrievals. 
If a correct image enters the top 10, it is termed a positive replacement effect; conversely, if an incorrect image enters, it is termed a negative replacement effect.
The replacement effect does not guarantee whether a correct or incorrect image will be replaced, and identifying the specific image replaced after using the combination method is challenging. 




Figure~\ref{fig:combination_effect} presents examples of positive and negative replacement effects for combinations of single caption from Internlm-xcomposer2-vl-7B and collective captions from LLaVA-NeXT-7B, as well as fine-grained and coarse-grained captions.

\noindent \textbf{Positive Replacement Effect Analysis}: In Figure~\ref{fig:combination_effect} (a), the correct label is topic 1. The single caption reads: ``A person is holding an ice cream cone in their hand.''. 
The collective caption reads: `` At 18:14, a person is seen holding a cone of ice cream, …, The scene is set on a beach, …''. While the single caption lacks beach context, it helps identify the specific image among collectively captioned images.
The image's ranking improved from 4,978th to 9th when combined with the collective caption score, originally ranked 5th. 

In Figure~\ref{fig:combination_effect} (b), the correct label is topic 2. The fine-grained caption states: ``Person at a restaurant table with a pint of Guinness and a glass of water. '' The coarse-grained caption states: ``Consuming a beverage in a restaurant with a pint of Guinness and empty dishes. ''
The coarse-grained caption suggests a restaurant setting, while the fine-grained caption confirms the specific image in the sequence. The ranking improved from 197th to 7th when combined with the coarse-grained caption's score, which initially ranked 4th.

\noindent \textbf{Negative Replacement Effect Analysis}: In Figure~\ref{fig:combination_effect} (c), the image is incorrectly categorized under topic 1. The single caption reads: ``The image captures a convenience store at 18:15, with the cash register and various snacks on display.'' The collective caption states: ``At 18:14, a person is seen holding a cone of ice cream, ready to enjoy a sweet treat. The scene is set on a beach.. ''. Although the sequence of images associated with the collective caption contains the correct labeled images, some accurate single captions within this sequence are not ranked higher. As a result, no improvement is observed after applying the combination method. The ranking improved from 2828th with the single caption to 4th after the combination, with the collective caption initially ranked 2nd.

In Figure~\ref{fig:combination_effect} (d), the image is incorrectly labeled as topic 10. The fine-grained caption is: ``Reading a text-heavy document on a brightly lit screen in darkness.'' The coarse-grained caption is: ``Reading and reviewing text-heavy documents with colorful highlights on a screen in darkness.'' These captions are similar due to the design of the merging method. 
The ranking difference between fine-grained and coarse-grained captions may be due to the GTE-large embedding model's performance or the lack of a suitable coarse-grained caption for Topic 10. 
The ranking increased from 153rd with the fine-grained caption to 7th after the combination, with the coarse-grained caption initially ranked 7th.

There are 41 positive replacement effects in the Internlm $\times$ Video-LLaVA combination and 32 in the GPT4 fine-grained $\times$ GPT4 coarse-grained combination. In contrast, there are 22 negative effects in the first combination and 24 in the second. While it remains unclear how many incorrect images are replaced by positive effects, these results provide insights into why the combination method surpasses single and collective caption methods with GTE-large text embeddings.

\begin{figure}[t]
  \centering
  \includegraphics[width=1\linewidth]{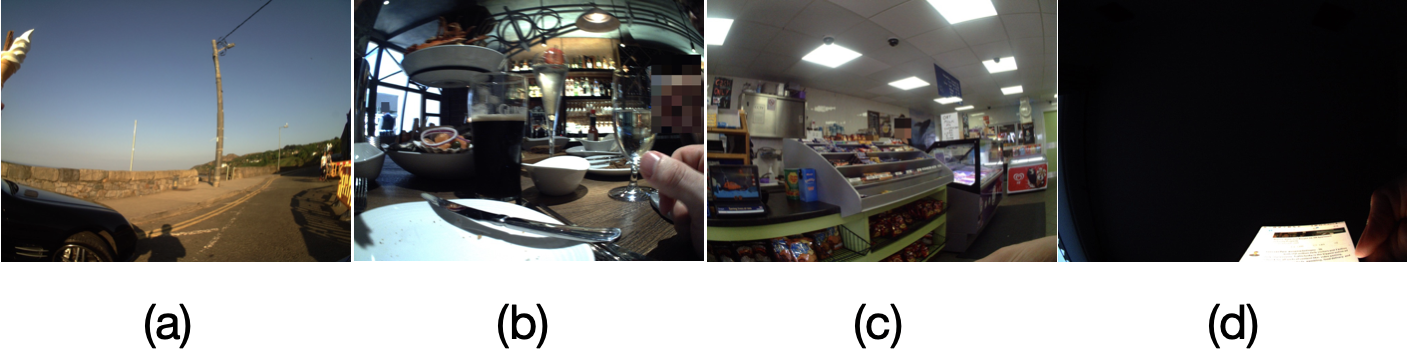}
  \caption{Examples of Combination effect.}
  \label{fig:combination_effect}
\end{figure}




\section{Conclusion and Future Work}
Recently, individuals increasingly document their lives through blogs, social media, photos, and videos, significantly contributing to the massive data generation. 
This vast amount of personal data holds potential for life assistance services, enabling users to retrieve related images from given text queries. 
However, the semantic gap between textual queries and visual lifelogs often impedes the retrieval of desired experiences.
To address this, we propose a Caption-Integrated Visual Lifelog Retrieval System that converts images into captions and uses a text embedding model to map both image captions and user queries into a unified vector space. Our approach outperforms baseline methods that directly embed images and text into the same vector space.

We also provide a textual dataset in our experiment generated from the NTCIR-14 Lifelog-3 dataset, to support further research in lifelogging retrieval and comprehension.
The dataset includes captions generated by single, collective, and merged caption methods, the latter offering fine-grained captions, group-level coarse-grained captions, and comprehensive experience summaries. This enables analysis of how LVLMs can reconstruct daily events through concatenated experience summaries or coarse-grained captions.

Future work will incorporate methods to improve cluster recall into our system, enhancing the diversity of retrieved images to better align with real-world user needs.
Furthermore, with advancements in LVLMs like GPT-4o and Gemini-Pro, we aim to explore how well-crafted prompts can reduce issues such as hallucinations and improve accuracy in identifying key elements within visual data.

\section*{Acknowledgements}
This research was partially supported by Google gift toward the work on Multimodal User Profiling for Personalized Information Access, National Science and Technology Council, Taiwan, under grants NSTC 112-2634-F-002-005-, and Ministry of Education (MOE) in Taiwan, under grants 113L900901.

\section*{IEEE Copyright Notice}
\noindent
\footnotesize
\textcopyright~2024 IEEE. Personal use of this material is permitted.
Permission from IEEE must be obtained for all other uses, in any current or future media,
including reprinting/republishing this material for advertising or promotional purposes,
creating new collective works, for resale or redistribution to servers or lists,
or reuse of any copyrighted component of this work in other works.\\
DOI: \href{https://ieeexplore.ieee.org/document/10825835}{10.1109/BigData62323.2024.10825835}

\bibliographystyle{IEEEtran}
\bibliography{IEEEabrv}

\begin{thebibliography}{10}
\providecommand{\url}[1]{#1}
\csname url@samestyle\endcsname
\providecommand{\newblock}{\relax}
\providecommand{\bibinfo}[2]{#2}
\providecommand{\BIBentrySTDinterwordspacing}{\spaceskip=0pt\relax}
\providecommand{\BIBentryALTinterwordstretchfactor}{4}
\providecommand{\BIBentryALTinterwordspacing}{\spaceskip=\fontdimen2\font plus
\BIBentryALTinterwordstretchfactor\fontdimen3\font minus \fontdimen4\font\relax}
\providecommand{\BIBforeignlanguage}[2]{{%
\expandafter\ifx\csname l@#1\endcsname\relax
\typeout{** WARNING: IEEEtran.bst: No hyphenation pattern has been}%
\typeout{** loaded for the language `#1'. Using the pattern for}%
\typeout{** the default language instead.}%
\else
\language=\csname l@#1\endcsname
\fi
#2}}
\providecommand{\BIBdecl}{\relax}
\BIBdecl

\bibitem{6563978}
T.~Maekawa, ``A sensor device for automatic food lifelogging that is embedded in home ceiling light: A preliminary investigation,'' in \emph{2013 7th International Conference on Pervasive Computing Technologies for Healthcare and Workshops}, 2013, pp. 405--407.

\bibitem{Yen_Huang_Chen_2021}
A.-Z. Yen, H.-H. Huang, and H.-H. Chen, ``Unanswerable question correction in question answering over personal knowledge base,'' \emph{AAAI}, vol.~35, no.~16, pp. 14\,266--14\,275, May 2021.

\bibitem{cad7ae14c7f047749df68f0a7f74a70b}
A.~Yen, M.~Fu, W.~Ang, T.~Chu, S.~Tsai, H.~Huang, and H.~Chen, ``\BIBforeignlanguage{English}{Visual lifelog retrieval: humans and machines interpretation on first-person images},'' \emph{\BIBforeignlanguage{English}{Multimedia Tools and Applications}}, vol.~82, no.~24, pp. 37\,757--37\,787, Oct. 2023.

\bibitem{chu2021vidlife}
T.-T. Chu, A.-Z. Yen, W.-H. Ang, H.-H. Huang, and H.-H. Chen, ``Vidlife: A dataset for life event extraction from videos,'' in \emph{CIKM}, 2021, pp. 4436--4444.

\bibitem{10.1145/3592573.3593101}
Q.-L. Tran, L.-D. Tran, B.~Nguyen, and C.~Gurrin, ``Memoriease: An interactive lifelog retrieval system for lsc’23,'' in \emph{Proceedings of the 6th Annual ACM Lifelog Search Challenge}, 2023, p. 30–35.

\bibitem{10.1145/3592573.3593105}
K.~Schoeffmann, ``lifexplore at the lifelog search challenge 2023,'' in \emph{Proceedings of the 6th Annual ACM Lifelog Search Challenge}, 2023, p. 53–58.

\bibitem{10.1145/3512729.3533006}
N.~Alam, Y.~Graham, and C.~Gurrin, ``Memento 2.0: An improved lifelog search engine for lsc'22,'' in \emph{Proceedings of the 5th Annual on Lifelog Search Challenge}, 2022, p. 2–7.

\bibitem{10.1145/3512729.3533012}
L.-D. Tran, M.-D. Nguyen, B.~Nguyen, H.~Lee, L.~Zhou, and C.~Gurrin, ``E-mysc\'{e}al: Embedding-based interactive lifelog retrieval system for lsc'22,'' in \emph{Proceedings of the 5th Annual on Lifelog Search Challenge}, 2022, p. 32–37.

\bibitem{10.1145/3592573.3593098}
T.-N. Nguyen, T.-K. Le, V.-T. Ninh, C.~Gurrin, M.-T. Tran, T.~B. Nguyen, G.~Healy, A.~Caputo, and S.~Smyth, ``E-lifeseeker: An interactive lifelog search engine for lsc’23,'' in \emph{Proceedings of the 6th Annual ACM Lifelog Search Challenge}, 2023, p. 13–17.

\bibitem{gurrin2019overview}
C.~Gurrin, H.~Joho, F.~Hopfgartner, L.~Zhou, V.-T. Ninh, T.-K. Le, R.~Albatal, D.-T. Dang-Nguyen, and G.~Healy, ``Overview of the ntcir-14 lifelog-3 task,'' in \emph{Proceedings of the 14th NTCIR conference}.\hskip 1em plus 0.5em minus 0.4em\relax NII, 2019, pp. 14--26.

\bibitem{Zhang_2021}
H.~Zhang, A.~Sun, W.~Jing, G.~Nan, L.~Zhen, J.~T. Zhou, and R.~S.~M. Goh, ``Video corpus moment retrieval with contrastive learning,'' in \emph{SIGIR}, ser. SIGIR ’21, 2021.

\bibitem{qasim2023densevideocaptioningsurvey}
I.~Qasim, A.~Horsch, and D.~K. Prasad, ``Dense video captioning: A survey of techniques, datasets and evaluation protocols,'' 2023.

\bibitem{zhou2024streamingdensevideocaptioning}
X.~Zhou, A.~Arnab, S.~Buch, S.~Yan, A.~Myers, X.~Xiong, A.~Nagrani, and C.~Schmid, ``Streaming dense video captioning,'' 2024.

\bibitem{openai2024gpt4}
J.~Achiam, S.~Adler, S.~Agarwal, L.~Ahmad, I.~Akkaya, F.~L. Aleman, D.~Almeida, J.~Altenschmidt, S.~Altman, S.~Anadkat \emph{et~al.}, ``Gpt-4 technical report,'' \emph{arXiv preprint arXiv:2303.08774}, 2023.

\bibitem{ImageCLEF19}
B.~Ionescu, H.~M\"uller, R.~P\'{e}teri, Y.~D. Cid, V.~Liauchuk, V.~Kovalev, D.~Klimuk, A.~Tarasau, A.~B. Abacha, S.~A. Hasan \emph{et~al.}, ``{ImageCLEF 2019}: Multimedia retrieval in medicine, lifelogging, security and nature,'' in \emph{Experimental IR Meets Multilinguality, Multimodality, and Interaction}, ser. Proceedings of the 10th International Conference of the CLEF Association (CLEF 2019), 2019.

\bibitem{radford2021learning}
A.~Radford, J.~W. Kim, C.~Hallacy, A.~Ramesh, G.~Goh, S.~Agarwal, G.~Sastry, A.~Askell, P.~Mishkin, J.~Clark, G.~Krueger, and I.~Sutskever, ``Learning transferable visual models from natural language supervision,'' 2021.

\bibitem{zhai2023sigmoid}
X.~Zhai, B.~Mustafa, A.~Kolesnikov, and L.~Beyer, ``Sigmoid loss for language image pre-training,'' 2023.

\bibitem{fang2023data}
A.~Fang, A.~M. Jose, A.~Jain, L.~Schmidt, A.~Toshev, and V.~Shankar, ``Data filtering networks,'' 2023.

\bibitem{laurençon2024matters}
H.~Laurençon, L.~Tronchon, M.~Cord, and V.~Sanh, ``What matters when building vision-language models?'' 2024.

\bibitem{liu2024llavanext}
H.~Liu, C.~Li, Y.~Li, B.~Li, Y.~Zhang, S.~Shen, and Y.~J. Lee, ``Llava-next: Improved reasoning, ocr, and world knowledge,'' January 2024.

\bibitem{dong2024internlmxcomposer2}
X.~Dong, P.~Zhang, Y.~Zang, Y.~Cao, B.~Wang, L.~Ouyang, X.~Wei, S.~Zhang, H.~Duan, M.~Cao, W.~Zhang, Y.~Li, H.~Yan, Y.~Gao, X.~Zhang, W.~Li, J.~Li, K.~Chen, C.~He, X.~Zhang, Y.~Qiao, D.~Lin, and J.~Wang, ``Internlm-xcomposer2: Mastering free-form text-image composition and comprehension in vision-language large model,'' 2024.

\bibitem{lin2023videollava}
B.~Lin, Y.~Ye, B.~Zhu, J.~Cui, M.~Ning, P.~Jin, and L.~Yuan, ``Video-llava: Learning united visual representation by alignment before projection,'' 2023.

\bibitem{liu2024world}
H.~Liu, W.~Yan, M.~Zaharia, and P.~Abbeel, ``World model on million-length video and language with blockwise ringattention,'' 2024.

\bibitem{oquab2024dinov2}
M.~Oquab, T.~Darcet, T.~Moutakanni, H.~Vo, M.~Szafraniec, V.~Khalidov, P.~Fernandez, D.~Haziza, F.~Massa, A.~El-Nouby, M.~Assran, N.~Ballas \emph{et~al.}, ``Dinov2: Learning robust visual features without supervision,'' 2024.

\bibitem{li2023general}
Z.~Li, X.~Zhang, Y.~Zhang, D.~Long, P.~Xie, and M.~Zhang, ``Towards general text embeddings with multi-stage contrastive learning,'' 2023.

\bibitem{chen2024bge}
J.~Chen, S.~Xiao, P.~Zhang, K.~Luo, D.~Lian, and Z.~Liu, ``Bge m3-embedding: Multi-lingual, multi-functionality, multi-granularity text embeddings through self-knowledge distillation,'' 2024.

\end{thebibliography}

\end{document}